\documentclass[prl,twocolumn,nofootinbib,showkeys,showpacs]{revtex4-1}
\usepackage{amssymb}
\usepackage{amsmath}
\usepackage{amstext}
\usepackage{amsfonts}
\usepackage{bbold}
\usepackage{bm}
\usepackage{graphics}
\usepackage{mathtools}
\usepackage{lmodern}
\usepackage{graphicx}

\usepackage[colorlinks=true,        
            linkcolor=blue,         
            citecolor=black,         
            urlcolor=black,          
            breaklinks=true]{hyperref}

\usepackage[capitalise]{cleveref}

\newcommand{\be}{\begin{equation}}
\newcommand{\ee}{\end{equation}}

\newcommand{\ti}{{\tt{ t}}}

\begin{document}

\title{On the fate of spacetime singularities}

\author{Federico Piazza}
\affiliation{Aix Marseille Univ, Universit\'e de Toulon, CNRS, CPT, Marseille, France }

\begin{abstract}
I investigate spacetime singularities from the point of view of the wavefunction of the universe. In order to extend the classical notion of geodesic incompleteness one has to include the proper time of an observer as a degree of freedom in the Wheeler DeWitt equation. This leads to a Schr\"odinger equation along the observer worldline. Near the singularity,  as in  the classical BLK treatment, I ignore spatial gradients and effectively describe the spacetime around the worldline in the mini-superspace approximation. Then the problem proves identical to a spherically symmetric scattering of a quantum particle off a central potential and singularity  avoidance is tantamount to unitary evolution for this system. Standard types of matter (dust, radiation) correspond to regular potentials and thus lead to a bounce. The most singular component, spatial anisotropy, is associated to a conserved charge and yields a negative inverse-square potential---like standard angular momentum, but with opposite sign. This potential is critical, in that the unitarity of the evolution depends on the actual numerical factor in front of it, i.e., on the anisotropy charge.

\end{abstract}

\maketitle


\section{Introduction}

 It is generally believed that physics beyond classical general relativity (GR) should make sense of spacetime singularities. Most commonly, a resolution is expected from the UV completion of the theory.  As we approach the singularity, higher curvature terms in the gravitational action become important and can be used as corrections in the classical equations of motion of the low-energy theory, providing hints for a slow-down of the collapse, or for a bounce.  Most pre-big bang~\cite{Veneziano:1991ek,Gasperini:1992em,Gasperini:2002bn} or bouncing models (\emph{e.g.}~\cite{Bojowald:2008zzb,Battefeld:2014uga,Agullo:2016tjh}) are essentially based on this idea. 
However, an alternative, low-energy path to the resolution of the singularity is also conceivable. 
 
 \vspace{.5em}
 
\noindent\emph{\textbf{The role of the Wheeler DeWitt equation }} --- One can attempt to ``go beyond" classical GR by considering its straightforward quantum mechanical formulation, \emph{i.e.}, the  Wheeler DeWitt (WdW) equation~\cite{DeWitt:1967yk}.
 In this approach, the central object is not a specific solution of the Einstein equations, but the gravitational wavefunction $\Psi$,  which can be used to obtain the  probability to find the metric and the other fields in some configuration. 
 Quantum mechanical systems are generally more stable than their classical counterparts, with eminent examples shaping the very birth of the quantum theory. But could we trust the WdW equation if it implies such important deviations from classical GR solutions?
  
  At first sight, the answer is \emph{no}.  In systems at weak-coupling, the path integral is well-approximated by a classical saddle point. (This should be contrasted, \emph{e.g.}, with QCD below the $\Lambda_{QCD}$ scale, where the path integral and the classical equations of motion clearly yield very different predictions.)    As a low-energy theory, GR only makes sense at weak-coupling, so one might be led to conclude that GR is inherently classical and that quantum effects are confined to the unknown UV completion.

However, this conclusion might be too pessimistic. As long as the gravitational wavefunction maintains its support on smooth sub-Planckian metrics it should be under the control of the low-energy theory, but can still display features  that cannot be reproduced by a point-like classical system. For example, quantum tunneling is made possible by the non-vanishing width of the wavefunction, which allows it to ``feel" regions of the potential away from its peak. In this case, a weak-coupling semiclassical treatment is indeed available, but it applies to the Euclidean  path integral, so it is not directly associated to the classical (Lorentzian) evolution.  In gravity, semiclassical techniques apparently oblivious to UV physics have greatly contributed to the understanding of black hole thermodynamics (\emph{e.g.}~\cite{Gibbons:1976ue,Hawking:1982dh,Almheiri:2020cfm} to cite a few).

An example more fitting to the present context is the no-boundary solution by Hartle and Hawking~\cite{Hartle:1983ai}, which satisfies WdW~\cite{Hartle:1983ai,Halliwell:1990qr} but shows departures from the classical Lorentzian behavior at early times~\cite{Lehners:2023yrj}. More generally, the semiclassical approximation breaks-down at the classical turning points (\emph{e.g.}~\cite{Landau:1991wop}, Sec. 46).

By reasoning along these lines, cosmic bounces have been proposed within the WdW framework that do not rely on higher curvature corrections~\cite{Gasperini:1996np,Gielen:2016fdb,Gasperini:2021eri,Gielen:2022dhg,Gielen:2022tzi,Lehners:2024qaw,Piazza:2025fbt,Sahota:2025tih}. Generally speaking, they are made possible  by the boundary conditions of the wavefunction---or---as I review shortly, by the Heisenberg uncertainty principle, which makes regular certain potentials that are classically singular.

 \vspace{.5em}

\noindent\emph{\textbf{The role of proper time ---}} The notion of proper time is central to the definition of a spacetime singularity. Classically, the latter is defined by geodesic incompleteness, \emph{i.e.} by the impossibility to extend certain timelike geodesics (or worldlines) past some values \emph{of their proper time}~\cite{Hawking:1973uf}.   This specification is crucial to characterize the singularity as ``something wrong happening a finite distance away".  For example,  by a coordinate transformation one can bring regular points at infinity to finite coordinate values (like in a Penrose diagram) without this implying the presence of a singularity. Or, real singularities may be concealed by a bad coordinate choice. 

This potential ambiguity is mirrored and made worse quantum mechanically. The WdW equation is famously timeless, with $\Psi$ depending only on \emph{the dynamical fields} $q^a$ of the theory. In order to interpret $\Psi(q^a)$ as an actual evolution, one is forced to a relational interpretation, in which one of the fields (say, $q^0$) is chosen to play the role of time (e.g.~\cite{Piazza:2024lpt}). One can then compute from $\Psi$ the probability for the remaining fields to be in some configuration ``at some time", \emph{i.e.}, conditional to a value of $q^0$.  By using different choices of relational time the behavior of the wavefunction close to the singularity has been studied in a number of works (e.g.~\cite{Gielen:2020abd,Gielen:2021igw,Hartnoll:2022snh}).

However, as emphasized above,  it should be the proper time of an observer, not some arbitrarily chosen field, that diagnoses a singularity.
As the WdW equation only deals with dynamical fields we should introduce the  trajectory of the observer and its proper time $\ti$ as new degrees of freedom. While we send the observer into the singularity we can compute $\Psi(q^a, \ti)$ along the worldline and obtain a probability conditional to the actual proper time $\ti$ of the observer. Equivalently,  we can compute the expectation values of the  $q^a$ variables \emph{at given} $\ti$. 
%
%
This seems the correct starting point to extend the notion of singularity---or avoidance thereof---into the quantum regime. Without incorporating the clock variable 
$\ti$, inspection of the ``standard" WdW wavefunction 
$\Psi(q^a)$ generally leads to inconclusive results. 

To clarify this, let me consider a very stupid example with only two fields, the scale factor $a$ and a scalar $\phi$. Say that $\Psi(a, \phi)$ turns out to be peaked along the field space trajectory $\phi = a$. Such a wavefunction has a clear relational interpretation~\cite{Piazza:2024lpt}. If, \emph{e.g.}, we use $a$ as ``time", we have, approximately, $\langle \phi \rangle = a$. Does this model contain a singularity or a bounce? 

Only when $\ti$ is added as a degree of freedom can we address this question. From $\Psi(a, \phi, \ti)$, one can compute the average values of $a$ and $\phi$ as functions of $\ti$. The probe should only slightly perturb the system, so $\Psi$ is still be peaked on a trajectory  belonging to the two-dimensional plane $a = \phi$. However, this can include very disparate situations. We could have, say,  $\langle a \rangle = \langle \phi \rangle = \ti^2 + 1$, which clearly represents a bounce. Or, we could find that both $\langle a \rangle$ and  $\langle \phi \rangle$ tend to infinity at $\ti = 0$---more precisely, that $\Psi$ does not evolve unitarily in $\ti$. In this case, we would simply conclude that the low-energy theory alone is not able to resolve the singularity.

\section{The Schr\"odinger equation}

Here I review how the inclusion of the proper time of the observer into the WdW formalism leads to a Schr\"odinger equation along the worldline~\cite{Piazza:2025fbt}. The use of the Schr\"odinger equation and/or proper time in quantum cosmology has  been advocated in a number of papers (e.g.~\cite{teitelboim1983proper,unruh1989time,Peter_2018,Malkiewicz:2022szx,Burns:2022fzs,Kaya_2023,Bergeron:2025eda}). 
Of particular relevance to the present work are Refs.~\cite{Vachaspati:2006ki,Greenwood:2008ht,Wang:2009ay,Saini:2014qpa} for application to singularities.

 \vspace{.5em}

\noindent\emph{\textbf{Quid est Tempus?}} --- Classical GR's answer to this question is unambiguous, at least \emph{locally}, \emph{i.e.} \emph{along an observer's worldline}. The invariant line element, or proper time, $d \ti$ is obtained by contracting the metric tensor $g_{\mu \nu}$ along the observer's trajectory---see eq.~\eqref{classicclock} below.  The actual  procedures that the observer needs to implement in order to measure proper time are not of concern here. By definition, a \emph{good clock} must tick homogeneously in the $\ti$ variable. 

When we upgrade to quantum gravity,  the observer's trajectory $x^\mu(\lambda)$  ($\lambda$ is a parameter along the trajectory) and its proper time $\ti$ should be promoted to quantum variables. The canonical pair $(\ti,p_\ti)$ lives on the worldline and is governed by the action
\begin{equation} \label{actionobs}
I_{\rm clock} = \int d\lambda\,
\left[
  p_\ti \,\frac{d\ti}{d \lambda}
  -
 \sqrt{- g_{\lambda \lambda} }\  p_\ti 
\right]\,.
\end{equation}
This should be supplemented by the action for the observer trajectory, which we can take as the standard geodesic one, $I_{\rm obs} = - m \int d\lambda \sqrt{-g_{\lambda \lambda}}$.

Variation with respect to $p_\ti$ gives the classical equation
\begin{equation} \label{classicclock}
\frac{d \ti}{d \lambda} = \sqrt{- g_{\lambda \lambda} } \equiv \sqrt{- g_{\mu \nu} \frac{d x^\mu}{d \lambda}\frac{d x^\nu}{d \lambda}}\, ,
\end{equation}
which coincides with the definition of proper time and thus motivates~\eqref{actionobs}.\footnote{By introducing an \emph{einbein} $e$ one can write~\eqref{actionobs} \emph{\`a la Polyakov} with $p_\ti$ appearing quadratically in the Hamiltonian, for example
\begin{equation} I_{\rm clock} = \int  d\lambda\,
\left[
  p_\ti \,\frac{d\ti}{d \lambda}
  -
 \sqrt{- g_{\lambda \lambda}}\ \left(\frac{p_\ti^2}{2 e} + \frac{e}{2}\right) 
\right]\,.
\end{equation} 
I thank G. Veneziano for pointing this out. This trick apparently allows one to write the action in standard ``Lagrangian" form, i.e., as a functional of $\ti(\lambda)$, which is not obvious from~\eqref{actionobs}.   }

Quantum mechanically, the variance associated with $\ti$ can be interpreted as the uncertainty to set the origin of our clock variable  along the worldline. Or, as Witten nicely puts it~\cite{Witten:2023xze},
\begin{quote}
\leftskip=6mm
\rightskip=6mm
\small When we take gravity to be dynamical, we have to take into account that the same observer worldline can be embedded in a given spacetime in different ways, differing by $\ti \rightarrow \ti \, +$ constant.
\end{quote}

 
\noindent\emph{\textbf{Synchronous gauge}} --- In this paper I adopt the \emph{synchronous gauge}, with the metric written as
\begin{equation} \label{metric}
ds^2 = - N^2 dt^2 + h_{ij} dx^i dx^j\, .
\end{equation}  
This gauge corresponds to fixing the spatial coordinates $x^i$ in such a way that the shifts $N_i = 0$. This can be done by identifying a congruence of geodesic observers, among which the observer provided with the clock~\eqref{classicclock} sits at $x^i = 0$.  The formal way to do so is to introduce a non-relativistic fluid type of matter, described by three scalar fields $\phi^I$ where $\phi^I = constant$ represent the trajectories of the volume elements of the fluid, or of the observers inside the congruence. By adopting the \emph{unitary gauge} choice $\phi^I = x^i$, the momentum constraint then implies $N_i = 0$ (e.g.~\cite{Nicolis:2022gzh}, App. D). The Hamiltonian constraint is the WdW equation, that I discuss shortly. 

This gauge has been used to study classical (spacelike) singularities by Belinsky, Lifshitz and Khalatnikov (BLK) in their seminal paper~\cite{Belinsky:1970ew}. One central hypothesis of that paper, validated by later numerical studies~\cite{Garfinkle:2003bb}, is that as we approach the singularity spatial gradients become irrelevant and $h_{ij}$ effectively becomes a function of time only. One heuristic way of seeing this is that the contribution of the gradients to the energy density scales as $a^{-4}$ for modes that are still inside the horizon. This should be confronted with the behavior of anisotropies that, classically, scale as $a^{-6}$. As we approach the singularity anisotropies tend to dominate the dynamics eventually, in the so-called \emph{Kasner phase}.

The full picture is in fact much richer. As the singularity is approached, certain directions of the spatial curvature begin to dominate over anisotropies, leading to a succession of increasingly shorter Kasner phases (\emph{i.e.} the so-called \emph{chaotic Mixmaster universe}~\cite{Misner:1969hg,Misner:1973prb,Montani:2007vu}).
In this paper, however, I restrict the analysis to a single Kasner phase and assume,  accordingly,  that both  $N$ and $h_{ij}$ in~\eqref{metric} are only time-dependent. In other words, I assume the mini-superspace approximation along the worldline.

 \vspace{.5em}

\noindent\emph{\textbf{L'uovo di Colombo}} --- The Einstein Hilbert action for this metric will be derived shortly but first notice that, in these coordinates, one can choose directly $\lambda = t$ in~\eqref{actionobs} and the Hamiltonian of the clock becomes 
\begin{equation}\label{intro_clock}
H_{\rm clock} = N p_\ti\, . 
\end{equation}
Because of time-reparametrization invariance, the Hamiltonian of gravity + matter also contains an overall factor of $N$. By including the clock  we thus have  
\begin{equation}
H_{\rm tot} = N {\cal H}_{\rm tot} = N(p_\ti + {\cal H})
\end{equation}
where $N {\cal H}$ is the Hamiltonian of the system \emph{without} the clock.
In the standard WdW approach one poses ${\cal H} \Psi(q^a) = 0$ ($q^a$ are now  mini-superspace quantum variables). By including the clock into the system, the WdW equation becomes 
${\cal H}_{\rm tot} \Psi(q^a, \ti) = 0$. Upon use of the canonical relation $p_\ti=-\,i \partial_\ti$ this is just the Schr\"odinger equation as advertized,
 \begin{equation}  \label{equation}
\setlength{\fboxsep}{3\fboxsep} \boxed{i \partial_\ti \Psi(q^a,\ti) = \mathcal{H}\,\Psi(q^a,\ti)\, ,}
 \end{equation} 
where  ${\cal H}$ is the Hamiltonian of the system in the absence of the clock. 

It looks now very natural to use $\ti$ as a conditional variable in the usual relational interpretation of the WdW wavefunction. The positive-definite conserved probability then reads
 \begin{equation} \label{probability}
dP(q^a|\ti) = \sqrt{-g} \, |\Psi(q^a,\ti)|^2 dq^a\, .
\end{equation} 

Notice that, contrary to the other---supposedly ``fundamental"---fields $q^a$, $\ti$ is merely a bookkeeper for all the microphysical phenomena that are practically used to measure time. 
 The fact that $H_{\rm clock}$ in~\eqref{intro_clock} is unbounded from below may reflect more the intrinsic limitations of a real physical clock, rather  than a problem with the model we are employing. As a matter of fact, the Schr\"odinger equation~\eqref{equation} can be made unstable only by (the non self-adjointness of-) the Hamiltonian operator 
${\cal H}$. Specifically, the kinetic term of the scale factor enters the latter with the ``wrong" sign, which represents the classical instability of the system against gravitational collapse. This instability, and its possible quantum resolution, is the whole focus of this paper. 

 \vspace{.5em}
 
\noindent\emph{\textbf{Time as a test field}} --- In~\cite{Piazza:2025fbt} a prescription for the ``correct" solutions of~\eqref{equation} was proposed, namely, that there is little or no-energy associated with the clock so that it represents a gentle probe of the system. Effectively, we are looking for solutions of~\eqref{equation} with total energy of  ${\cal O}(\alpha)$, with $\alpha$ the small gravitational coupling introduced in the next section (see~\cite{Piazza:2025fbt} for more details).

 \section{Approaching the singularity}

As already stated, I will make the hypothesis that the lapse function $N$ and the spatial metric components $h_{ij}$ in~\eqref{metric} depend only on time. Under this assumption, the Einstein Hilbert action reads  
\begin{equation} \label{bianchiaction}
I_{\rm EH} = \frac{L^3}{16 \pi G_N}\int  \frac{dt }{4 N} \sqrt{h } \left(h^{il} h^{jm} - h^{ij} h^{lm}\right) \dot{h}_{ij} \dot{h}_{lm} \, ,
\end{equation}
where the spatial integration of the action has already been performed and has produced the spatial volume $L^3$. This can be thought of as the volume of the spatial section of the comoving ``tube" of spacetime around the observer at the time when the scale factor evaluates one.

 \vspace{.5em}

\noindent\emph{\textbf{A convenient parameterization}} --- In the absence of anisotropic stresses the metric $h_{ij}$  can be conveniently parametrized as (e.g.~\cite{Nicolis:2022gzh})
\begin{equation}
h_{ij} = a^2 \begin{pmatrix}
e^{4 (\beta_+ + \sqrt{3} \beta_-)/3} &0&0\\
0&e^{4 (\beta_+ - \sqrt{3} \beta_-)/3}&0\\
0&0&e^{-8 \beta_+/3} 
\end{pmatrix}\, .
\end{equation}
In this parameterization the action~\eqref{bianchiaction} reads
\begin{equation} \label{bianchiaction2}
I_{\rm EH} = \frac{3 L^3}{8 \pi G_N}\int  \frac{dt }{N} a^3 \left[- \frac{\dot a^2}{a^2} + \frac49 \left(\dot \beta_+^2  + \dot \beta_-^2 \right) \right]  \, .
\end{equation}

As expected, the anisotropies $\beta_+$ and $\beta_-$ behave as shift-symmetric scalar fields and can thus be associated to conserved charges. It is convenient to define a rescaled scale factor 
\begin{equation}
r = a^{3/2}\, ,
\end{equation}
which appears in the action with a standard kinetic term, 
\begin{equation} \label{bianchiaction3}
I_{\rm EH} = \frac{L}{2 \alpha}\int  \frac{dt }{N} \left[- \dot r^2 +  r^2\left(\dot \beta_+^2  + \dot \beta_-^2 \right)\right]  \, .
\end{equation}

The above action should be compared to the kinetic term of a particle in spherical coordinates, with
$
\vec v\, ^2 = \dot r^2 + r^2(\dot \theta^2 + \sin^2 \theta \, \dot \varphi^2)\, .
$
The crucial sign difference gives rise to a centripetal (rather than centrifugal-) force. In this respect, $r$ is by all means a ``radial" coordinate. 
In the above, I have introduced the small gravitational coupling 
\begin{equation}
\alpha = \frac{3 \pi G_N}{L^2}\, ,
\end{equation}
associated with the comoving region of space that we are considering.
The conserved momenta conjugate to the anisotropies read
\begin{equation} \label{charges}
p_\pm = \frac{\partial \cal L}{\,\partial \dot \beta_\pm} = \frac{r^2 L}{ \alpha}  \dot \beta_\pm\, .
\end{equation}

In this mini-superspace limit, a perfect fluid of equation of state $w = p/\rho$ can be modeled by the matter action (\emph{e.g.}~\cite{Nicolis:2022gzh})
\begin{equation} \label{matter}
I_{\rm matter} = - \frac{b}{ \alpha L}\int dt N \, r^{-2 w}, 
\end{equation}
where the number $b$ quantifies how much matter there is at around $r \simeq 1$. 
We see that this type of matter contributes a spherically symmetric potential to the total action of the system. 

 \vspace{.5em}

\noindent\emph{\textbf{Like an s-wave scattering}} --- The kinetic part of the action~\eqref{bianchiaction3} defines a metric in field space. The Hamiltonian operator should be built with the corresponding covariant Laplacian~\cite{Nicolis:2022gzh}. This gives the time-dependent Schr\"odinger equation
\begin{equation}
i \alpha L \partial_\ti \Psi = 
\left[\frac{\alpha^2}{2}\left(\partial_r^2 + \frac{2}{r} \partial_r - \frac{1}{ r^2}\, \Delta^{\!(2)} \right) + b r^{-2 w} \right] \Psi \, ,
\end{equation}   
where $\Delta^{\!(2)} = \partial_+^2 + \partial_-^2$ is the Laplacian in the flat two-dimensional anisotropy space and, as opposed to the schematic~\eqref{equation}, all factors of $\alpha$ and $L$ have been made explicit. However, by measuring time in units of $L$, we can effectively set $L=1$ from now on.

Anisotropies clearly play the role of the conserved angular momentum, with the difference that they add instability to the system, because they act as a centripetal force. 
 Anisotropy eigenfunctions, the analogous of spherical harmonics, can be chosen as 
$ Z_{p_+,p_-} = e^{i(p_+ \beta_+ + p_- \beta_-)}$.

They satisfy 
\begin{equation}
- \Delta^{\!(2)} Z_{p_+, p_-}(\beta_+, \beta_-) = (p_+^2 + p_-^2) Z_{p_+, p_-}(\beta_+, \beta_-) \, .
 \end{equation}
It is useful to define the (positive) \emph{total anisotropy charge}, 
\begin{equation}\label{charge}
Q \equiv \frac{\alpha^2}{2}\left(p_+^2 + p_-^2\right) \, .
\end{equation}

 By standard separation of variables, we can enquire about the time evolution of a state of given $Q$, 
 \begin{equation}
 \Psi(r, \beta_+, \beta_-, \ti) = R(r,\ti; Q) \, Z_{p_+,p_-}(\beta_+, \beta_-)\, ,
 \end{equation}
 where~\eqref{charge} is meant to apply.  $R$ is the analogous of the radial wavefunction in a central potential and satisfies the radial time-dependent Schr\"odinger equation
 \begin{equation}
\setlength{\fboxsep}{3\fboxsep} \boxed{i \alpha \, \partial_\ti R = 
\left[\frac{\alpha^2}{2}\left(\partial_r^2 + \frac{2}{r} \partial_r\right)  + \frac{Q}{r^2}  + \frac{b}{ r^{2 w}} \right] R \, . }
 \end{equation}

If we perform a time reversal on the above equation, $\ti \rightarrow - \ti$, we are effectively switching the sign of the RHS. The kinetic term becomes the standard one (``with a minus sign") and the effective potential for this system reads
\begin{equation}\label{veff}
V_{\rm eff}(r) =  - \frac{Q}{r^2}  - \frac{b}{ r^{2 w}}\, .
 \end{equation}

As predicted, anisotropies contribute a \emph{negative} effective potential and act as a centripetal force. This is analogous to having $\ell (\ell+1)<0$ in the Hydrogen atom. I now review what is the maximally singular behavior for the potential to be compatible with unitary evolution. I will show that typical matter ($w = 0$ for non-relativistic matter, $w = 1/3$ for radiation) contributes to the potential in a regular way and that the only possible source of instability is given by anisotropy, depending on its  actual amount, \emph{i.e.} on the value of $Q$. 
 
 \section{Unitarity}
 
 Consider the following Hamiltonian of a particle, in $d$ spatial dimensions for generality,
 \begin{equation} \label{hamiltonian}
 H = \vec p^{\; 2} - \frac{\gamma}{r^{\beta}}\, ,
 \end{equation}
  with $\gamma \geq 0$ and $\beta \geq 0$. Classically, this system is unstable because the Hamiltonian is unbounded from below. 
  This allows potential energy to be converted indefinitely into kinetic energy, leading to runaway singular solutions, as in the case of a radial infall toward the origin.
Quantum mechanically, however, the system~\eqref{hamiltonian} is \emph{regular} if~\cite{Landau:1991wop} 
\begin{equation} \label{stability}
\beta <2\, , \ \ \qquad  {\rm or} \ \, \ \qquad \beta = 2 \, , \  \gamma < \frac{(d-2)^2}4\, .
\end{equation}
Part of this result is a simple consequence of the Heisenberg uncertainty principle. By applying $\Delta r \Delta p >1/2$, one finds the following lower bound to~\eqref{hamiltonian},
  \begin{equation} 
  H> \frac{1}{4 r^2} - \frac{\gamma}{r^{\beta}}\, .
  \end{equation}
The RHS of the above inequality is itself a function bounded from below if $\beta<2$, or if $\gamma<1/4$ in the marginal case $\beta =2$. Remarkably, this simple heuristic argument reproduces the correct power-law behavior of the potential and, in $d=3$, also the critical numerical value of $\gamma$ as in~\eqref{stability}. According to the classical analysis of~\cite{Landau:1991wop}  a potential more singular than~\eqref{stability} simply leads to the infall of the quantum particle into the center. 

 \vspace{.5em}

\noindent\emph{\textbf{Close to $r=0$}} --- A more quantitative way to see~\eqref{stability} is the following. Let us consider the eigenvalue problem associated to the Hamiltonian~\eqref{hamiltonian},
\begin{equation}  \label{eigen}
\left[- \frac{1}{r^{d-1}} \frac{d}{dr}\left(r^{d-1} \frac{d}{dr}\right) - \frac{\gamma}{r^{\beta}}\right]R(r) = E R(r)\, .
\end{equation}
In order for the probability~\eqref{probability} to be finite close to the origin,  one needs to have 
\begin{equation} \label{norma}
R\sim r^{\lambda} \qquad {\rm with} \qquad \lambda > - d/2.
\end{equation}

On the other hand, for $R \sim r^\lambda$ equation~\eqref{eigen} gives
\begin{equation} \label{expansion}
-\lambda (\lambda + d -2 )\, r^{\lambda -2} - \gamma \, r^{\lambda - \beta} = E r^\lambda\, .
\end{equation}

Let us consider the case $\beta<2$ first. In this case the first term on the LHS of~\eqref{expansion} is the most singular and one needs $\lambda = 0$ or $\lambda = 2 - d$ to get rid of it. The correct choice is the less singular, $\lambda = 0$, which is also the leading behavior of the wavefunction close to the origin in the hydrogen atom at $\ell =0$. This behavior is also largely compatible with the normalization condition~\eqref{norma}. 
The second term on the LHS of~\eqref{expansion} is then taken care of by the subleading piece of the expansion. 

When $\beta>2$ the second term on the LHS of~\eqref{expansion} dominates, and it cannot be set to zero or compensated in any obvious way. What happens in this case is that the phase of the wavefunction increases indefinitely while approaching the singularity. This can be seen~\cite{Perelomov:1970fz} already at leading order in the semi-classical expansion $R = e^{i (S_0/\hbar + S_1 + \dots)}$, which gives
\begin{equation}
S'_0(r) \sim \pm \, \, r^{-\beta/2}\, .
\end{equation}
 
 This increasingly oscillating phase makes it impossible to extract the phase-shift of the scattering wavefunction~\cite{Coon:2002sua}, with different regularizations/prescriptions yielding different results for the $S$-matrix elements. Another way of putting it is that the Hamiltonian operator needs a self-adjoint extension for $\beta>2$~\cite{Bawin:2003dm}, and such a procedure is not unique. 
 
In the critical case $\beta = 2$~\cite{Landau:1991wop} the two terms on the LHS of~\eqref{expansion} are equally important so one needs to impose $\lambda (\lambda +d-2 ) + \gamma =0$, or  
\begin{equation} \label{roots}
\lambda = 1 -\frac d2 \pm \sqrt{\frac{(d-2)^2}{4} - \gamma}\, .
  \end{equation}
  When $\gamma<\frac{(d-2)^2}{4}$ these exponents are real and negative and we are thus in a case essentially identical to $\beta<2$. However, for $\gamma >\frac{(d-2)^2}{4}$ the roots~\eqref{roots} develop an imaginary part and we find ourselves in a case analogous to $\beta>2$, with a rapidly oscillating phase close to the origin that does not allow a definite calculation of the phase shift. In $d=3$, for  $\gamma> \gamma_{\rm crit} = 1/4$,
    \begin{equation}
   R  \sim r^{-1/2} \sin\left( \delta \ln r + \theta\right) 
   \end{equation}
 with $\delta = \sqrt{\gamma - 1/4}$ and for some unknown phase $\theta$. In this critical case the phase oscillation is logarithmic in $r$, instead of  power-law. 
  
  \section{Discussion}

Classically, the energy density associated with spatial anisotropy scales as $a^{-6}$. Among all the components commonly considered (e.g. dust: $a^{-3}$, radiation: $a^{-4}$) this is the most singular behavior in the approach to a singularity. However, as shown such a behavior is still marginally bearable quantum mechanically.  In the presence of anisotropy the wavefunction of the universe is separable into subspaces of given anisotropy charges, much like the partial wave expansion into angular momentum components. 
This charge multiplies the most singular part of the effective potential~\eqref{veff}, which corresponds to the critical case $\beta = 2$. As shown in the last section, in this case, the stability of the system  depends on the numerical coefficient $\gamma$, or on the total anisotropy charge $Q$. 
The fate of our observer thus depends on the amount of anisotropy in its surroundings. However, when  the correct units are re-established, the catastrophe seems inevitable. From the previous analysis, the critical value of the anisotropy charge $Q$ is $\alpha^2$-suppressed, 
\begin{equation}
Q_{\rm crit} = \frac{\alpha^2}{8} \, .
\end{equation}
It would seems that a regular bounce would require to prepare the system in an almost perfectly isotropic state.

My analysis trivially extends to possible additional scalar fields present in the theory, whose kinetic terms would generically tend to dominate over the potential terms in the approach to the singularity, and behave like anisotropy. In this case the scattering problem becomes $d = 3+n$ dimensional, with $n$ the number of additional fields and the critical value of $\gamma$ given in general by~\eqref{stability}.

In realistic situations the system might undergo a long period of radiation domination before anisotropies become relevant.  If the transition between the $r^{-2/3}$ behavior of radiation domination and the $r^{-2}$ of anisotropy in~\eqref{veff} happens, say,  at $r_0$, one might hope that anisotropies be effectively irrelevant if the width of the wavepacket is much larger than $r_0$. 
At the same time, I have just studied the Kasner phase here---it will be interesting to re-introduce the effects of spatial curvature and study the quantum version of the chaotic Mixmaster universe~\cite{Belinsky:1970ew,Misner:1969hg,Misner:1973prb,Montani:2007vu}. While classically the singularity is still reached after a series of increasingly shorter Kasner phases, the Heisenberg principle could definitely help stabilizing the system in this case. Finally, it would be very nice to be comforted in these results by a dual holographic description extending well inside the (AdS-) black hole interior, along the lines of e.g.~\cite{Hartnoll:2022snh,DeClerck:2023fax}.

\vspace{1em}

\begin{acknowledgments}
\noindent\textbf{\emph{Acknowledgments}} --- I acknowledge inspiring  exchanges with Jose Beltr\'an Jim\'enez, Maurizio Gasperini, Steffen Gielen,  Alberto Nicolis,  Alessandro Podo, Riccardo Rattazzi,  Andrew Tolley and Sim\'eon Vareilles. I am particularly indebted to Mehrdad Mirbabayi and Gabriele Veneziano for important remarks on the draft. This work received support from the French government under the France 2030 investment plan, as part of the Initiative d'Excellence d'Aix-Marseille Universit\'e - A*MIDEX (AMX-19-IET-012). It was also supported by the ``action th\'ematique" Cosmology-Galaxies (ATCG) of the CNRS/INSU PN Astro and by the {\it Agence Nationale de la Recherche} under the grant ANR-24-CE31-6963-01.
\end{acknowledgments}

\bibliographystyle{apsrev4-1-noclass}
\bibliography{references}

\end{document}